\documentclass[letterpaper,10.5 pt]{article}   
\usepackage{osajnl2} 
\usepackage{cite}
\usepackage{amsmath}
\usepackage{graphicx}
\usepackage{url}
\usepackage{subfigure}
\usepackage{color}
\usepackage{floatrow}
\usepackage{multirow}

\newcommand{\eqn}[1]{Eq.(\ref{#1})}

\newcommand{\Eqn}[1]{Equation(\ref{#1})}

\begin{document}

\title{Generalized Nonlinear Wave Equation in Frequency Domain}


\author{Hairun Guo${^*}$, Xianglong Zeng and Morten Bache}
\address{Group of ultrafast nonlinear optics, Department of photonics engineering (DTU Fotonik), Technical University of Denmark, DK-2800, Kgs. Lyngby, Denmark}
\address{$^*$Corresponding author: haig@fotonik.dtu.dk}

\begin{abstract}
  We interpret the forward Maxwell equation with up to third order induced polarizations and get so called nonlinear wave equation in frequency domain (NWEF), which is based on Maxwell wave equation and using slowly varying spectral amplitude approximation. The NWEF is generalized in concept as it directly describes the electric field dynamics rather than the envelope dynamics and because it concludes most current-interested nonlinear processes such as three-wave mixing, four-wave-mixing and material Raman effects. We give two sets of NWEF, one is a 1+1D equation describing the (approximated) planar wave propagation in nonlinear bulk material and the other corresponds to the propagation in a waveguide structure.
\end{abstract}


\maketitle 

\section*{Introduction}
\label{sec-1}
There are three parts in the derivation. First, we review the derivation of Maxwell's wave equation in frequency domain and the expression of nonlinear induced polarization. Second, in the approximation of planar wave propagation, 1+1D nonlinear wave equation in frequency domain (NWEF) is derived as an interpretation of the 1+1D forward Maxwell equation with up to third order induced polarizations. Third, considering a waveguide structure, NWEF dependent on spatial mode profile is derived.

\section{Maxwell's wave equation and nonlinear induced polarization}
\label{sec-2}
We start from Maxwell's equations and material equations shown below \cite{yariv2006photonics,boyd2008nonlinear,agrawal2012nonlinear}:

\begin{equation}
\nabla  \times {\bf{E}} =  - \frac{{\partial {\bf{B}}}}{{\partial t}} ,
\nabla  \times {\bf{H}} =  - \frac{{\partial {\bf{D}}}}{{\partial t}} + {\bf{J}}
\label{Eq-Maxwell-1}
\end{equation}

\begin{equation}
\nabla  \bullet {\bf{B}} = 0 ,
\nabla  \bullet {\bf{D}} = \rho ,
\nabla  \bullet {\bf{J}} =  - \frac{{\partial \rho }}{{\partial t}}
\label{Eq-Maxwell-2}
\end{equation}

\begin{equation}
{\bf{D}} = {\varepsilon _0}{\bf{E}} + {\bf{P}} ,
{\bf{B}} = {\mu _0}{\bf{H}} + {\bf{M}} ,
{\bf{J}} = \sigma {\bf{E}}
\label{Eq-Material}
\end{equation}
"${\nabla }$" is the Laplace operator, ${{\bf{E}}}$ and ${{\bf{H}}}$ indicate the electric field (unit: ${\textstyle{V\over m}}$) and magnetic field (unit: ${\textstyle{A\over m}}$) vectors, ${{\bf{D}}}$ and ${{\bf{B}}}$ indicate the electric and magnetic flux densities. ${{\bf{J}}}$ is the current density vector and ${\rho }$ is the charge density, representing the sources for the electromagnetic field. ${\varepsilon _0}$ is the vacuum permittivity (unit: ${\textstyle{F\over m}}$) and ${{\mu _0}}$ is the vacuum permeability (unit: ${\textstyle{H\over m}}$). ${{\bf{P}}}$ and ${{\bf{M}}}$ indicate the induced electric and magnetic polarization.

In the absence of free charges in a nonmagnetic medium, we have ${{\bf{M}} = 0}$, ${{\bf{J}} = 0}$ and ${\rho  = 0}$. Hence the Maxwell's equations is simplified to:

\begin{equation}
\nabla  \times {\bf{E}} =  - \frac{{\partial {\bf{B}}}}{{\partial t}} ,
\nabla  \times {\bf{H}} = \frac{{\partial {\bf{D}}}}{{\partial t}}
\label{Eq-Maxwell-simplified}
\end{equation}

After Fourier transforming the above equations (${\tilde F(\omega ) = \int_{ - \infty }^\infty  {F(t){e^{ - i\omega t}}} dt}$ , matching the DFT algorithm in @MATLAB), we have:
\begin{equation}
\nabla  \times {\bf{\tilde E}} =  - i\omega {\bf{\tilde B}} =  - i\omega {\mu _0}{\bf{\tilde H}} ,
\nabla  \times {\bf{\tilde H}} = i\omega {\bf{\tilde D}} = i\omega ({\varepsilon _0}{\bf{\tilde E}} + {\bf{\tilde P}})
\label{Eq-Maxwell-simplified-FD}
\end{equation}

To derive the wave equation regarding the electric field, we use the relation:
\begin{equation}
\nabla  \times (\nabla  \times {\bf{\tilde E}}) = \nabla (\nabla  \bullet {\bf{\tilde E}}) - {\nabla ^2}{\bf{\tilde E}}
\label{Eq-relation}
\end{equation}
in which the left side can be extended as:
\begin{multline}
\nabla  \times (\nabla  \times {\bf{\tilde E}})
= \nabla  \times ( - i\omega {\mu _0}{\bf{\tilde H}})
=  - i\omega {\mu _0} \cdot (\nabla  \times {\bf{\tilde H}}) \\
=  - i\omega {\mu _0} \cdot i\omega ({\varepsilon _0}{\bf{\tilde E}} + {\bf{\tilde P}})
= {\omega ^2}{\mu _0}{\varepsilon _0}{\bf{\tilde E}} + {\omega ^2}{\mu _0}{\bf{\tilde P}}
= k_0^2(\omega ){\bf{\tilde E}} + {\omega ^2}{\mu _0}{\bf{\tilde P}}
\label{Eq-Maxwell-simplified-FD-left}
\end{multline}

On the right side, we have:
\begin{equation}
\nabla (\nabla  \bullet {\bf{\tilde E}}) - {\nabla ^2}{\bf{\tilde E}} \approx  - {\nabla ^2}{\bf{\tilde E}}
\label{Eq-Maxwell-simplified-FD-right}
\end{equation}
which is supported if: 1) the high order induced electric polarizations (nonlinear induced polarizations) are considered perturbations to the first order induced polarization (linear induced polarization), i.e. ${\bf{\tilde P}} = {{\bf{\tilde P}}_L} + {{\bf{\tilde P}}_{NL}} \approx {{\bf{\tilde P}}_L} = 2\pi {\varepsilon _0}{\tilde \chi ^{(1)}}{\bf{\tilde E}}$; 2) the relative permittivity ${{\varepsilon _r} = 1 + 2\pi {\tilde \chi ^{(1)}}}$ is independent on the spatial distribution, i.e. ${\nabla  \bullet {\bf{\tilde E}} = \nabla  \bullet {\bf{\tilde D}}/{\varepsilon _0}{\varepsilon _r} = 0}$.

Therefore, in frequency domain, we have the Maxwell's wave equation for the electric field:
\begin{equation}
{\nabla ^2}{\bf{\tilde E}} + k_0^2(\omega ){\bf{\tilde E}} + {\omega ^2}{\mu _0}{\bf{\tilde P}} = 0
\label{Eq-WaveEq-FD}
\end{equation}
In time-domain, it is:
\begin{equation}
{\nabla ^2}{\bf{E}} - \frac{1}{{{c^2}}}\frac{{{\partial ^2}{\bf{E}}}}{{\partial {t^2}}} - {\mu _0}\frac{{{\partial ^2}{\bf{P}}}}{{\partial {t^2}}} = 0
\label{Eq-WaveEq}
\end{equation}

The induced polarization is further expressed as:
\begin{equation}
{\bf{P}} = {{\bf{P}}^{(1)}} + {{\bf{P}}^{(2)}} + {{\bf{P}}^{(3)}} +  \cdots  + {{\bf{P}}^{(m)}}
\label{Eq-P}
\end{equation}
where the generalized expression of both the linear induced polarization ${{{\bf{P}}^{(1)}} = {{\bf{P}}_L}}$ and the nonlinear induced polarization ${{{\bf{P}}^{(m)}}}$ is:
\begin{multline}
{{\bf{P}}^{(m)}}
= {\varepsilon _0}\int_{ - \infty }^\infty  {d{t_1}} \int_{ - \infty }^\infty  {d{t_2}}  \cdots \int_{ - \infty }^\infty  {d{t_m}} {\chi ^{(m)}}({t_1},{t_2}, \cdots ,{t_m})|{\bf{E}}(t - {t_1}){\bf{E}}(t - {t_2}) \cdots {\bf{E}}(t - {t_m})
\\
= {\varepsilon _0}\int_{ - \infty }^\infty  {d{t_1}} \int_{ - \infty }^\infty  {d{t_2}}  \cdots \int_{ - \infty }^\infty  {d{t_m}} {\chi ^{(m)}}(t - {t_1},t - {t_2}, \cdots ,t - {t_m})|{\bf{E}}({t_1}){\bf{E}}({t_2}) \cdots {\bf{E}}({t_m})
\\
= {\varepsilon _0}\int_{ - \infty }^\infty  {d{\omega _1}} \int_{ - \infty }^\infty  {d{\omega _2}}  \cdots \int_{ - \infty }^\infty  {d{\omega _m}{{\tilde \chi }^{(m)}}({\omega _1},{\omega _2}, \cdots ,{\omega _m})|{\bf{\tilde E}}({\omega _1}){\bf{\tilde E}}({\omega _2}) \cdots {\bf{\tilde E}}({\omega _m}){e^{it\sum {{\omega _i}} }}}
\label{Eq-p_m}
\end{multline}
where:
\begin{equation}
{\tilde \chi ^{(m)}}({\omega _1},{\omega _2}, \cdots {\omega _m}) = \frac{1}{{{{\left( {2\pi } \right)}^m}}}\int_{ - \infty }^\infty  {d{t_1}} \int_{ - \infty }^\infty  {d{t_2}}  \cdots \int_{ - \infty }^\infty  {d{t_m}{\chi ^{(m)}}({t_1},{t_2}, \cdots {t_m}){e^{ - i\sum {{\omega _i}{t_i}} }}}
\label{Eq-chi_m}
\end{equation}

${{\chi ^{(m)}}}$ is the temporal response function of the material, also called susceptibility in frequency domain, which is a (m+1)-rank tensor. The calculations among the electric fields are dyadic product which result in an m-rank tensor. "${|}$" indicate the multiple tensor product between two tensors, i.e. a (m+1)-rank tensor and a m-rank tensor (or dyadic tensor). Therefore, the induced polarization ${{{\bf{P}}^{(m)}}}$ is a vector. In frequency domain, it has:
\begin{multline}
{{\bf{\tilde P}}^{(m)}}(\omega )
= \int_{ - \infty }^{ + \infty } {{{\bf{P}}^{(m)}}{e^{ - i\omega t}}dt}
\\
= {\varepsilon _0}\int_{ - \infty }^\infty  {d{\omega _1}} \cdots \int_{ - \infty }^\infty  {d{\omega _m}{{\tilde \chi }^{(m)}}({\omega _1}, \cdots ,{\omega _m})|{\bf{\tilde E}}({\omega _1}) \cdots {\bf{\tilde E}}({\omega _m})} \int_{ - \infty }^{ + \infty } {{e^{ - i\left( {\omega  - \sum {{\omega _i}} } \right)t}}dt}
\\
= 2\pi {\varepsilon _0}\int_{ - \infty }^\infty  {d{\omega _1}} \cdots \int_{ - \infty }^\infty  {d{\omega _m}{{\tilde \chi }^{(m)}}({\omega _1}, \cdots ,{\omega _m})|{\bf{\tilde E}}({\omega _1}) \cdots {\bf{\tilde E}}({\omega _m})\delta (\omega  - \sum {{\omega _i}} )}
\label{Eq-p_m-FD-1}
\end{multline}

Here, the delta function implies that the induced polarization always corresponds to the frequency which equals to the sum of the frequencies of the contributing electric fields.

If setting ${\Omega  = \sum {{\omega _i}} }$, i.e. ${d\Omega  = d{\omega _i}}$ and ${\int_{ - \infty }^{ + \infty } {\delta (\omega  - \Omega )d\Omega }  = 1}$, the above equation becomes:
\begin{multline}
{{\bf{\tilde P}}^{(m)}}(\Omega ) = 2\pi {\varepsilon _0}\int_{ - \infty }^\infty  {d{\omega _1}} \int_{ - \infty }^\infty  {d{\omega _2}}  \cdots \int_{ - \infty }^\infty  {d{\omega _{m - 1}}}
\\
\times {\tilde \chi ^{(m)}}({\omega _1},{\omega _2}, \cdots ,\Omega  - \sum\nolimits_1^{m - 1} {{\omega _i}} )|{\bf{E}}({\omega _1}){\bf{E}}({\omega _2}) \cdots {\bf{E}}(\Omega  - \sum\nolimits_1^{m - 1} {{\omega _i}} ),  {m \ge 2}
\label{Eq-p_m-FD-2}
\end{multline}
and ${{{{\bf{\tilde P}}}^{(1)}}(\Omega ) = 2\pi {\varepsilon _0}{{\tilde \chi }^{(1)}}(\Omega ) \bullet {\bf{\tilde E}}(\Omega )}$.

Since ${{{\bf{P}}^{(m)}}}$ is a vector, it can be written as a sum of its components, each casting to one dimension, i.e. ${{{\bf{P}}^{(m)}} = \sum {\hat jP_j^{(m)}} }$. Analogously, The nonlinear response tensor ${{\chi ^{(m)}}}$ has ${\chi ^{(m)}} = \sum {\hat j{\bf{R}}_j^{(m)}} $. Now ${{\bf{R}}_j^{(m)}}$ is an m-rank tensor and has ${{\bf{R}}_j^{(m)} = \sum\limits_{{\alpha _1} \cdots {\alpha _m}} {\left[ {\left( {\mathop \Pi \limits_{s = 1}^m {{\hat \alpha }_s}} \right) \cdot \chi _{j;{\alpha _1} \cdots {\alpha _m}}^{(m)}} \right]}}$, where ${j, {\alpha _1}, \cdots ,{\alpha _m}}$ is dimension mark.

Then, the component of the induced polarization ${P_j^{(m)}}$ is:
\begin{multline}
P_j^{(m)} (t)
= {\varepsilon _0}\int_{ - \infty }^\infty  {d{t_1}} \int_{ - \infty }^\infty  {d{t_2}}  \cdots \int_{ - \infty }^\infty  {d{t_m}{\bf{R}}_j^{(m)}(t - {t_1},t - {t_2}, \cdots ,t - {t_m})|{\bf{E}}({t_1}){\bf{E}}({t_2}) \cdots {\bf{E}}({t_m})}
\\
= {\varepsilon _0}\sum\limits_{{\alpha _1} \cdots {\alpha _m}} {\left\{ {\int_{ - \infty }^\infty  {d{t_1}}  \cdots \int_{ - \infty }^\infty  {d{t_m}\chi _{j;{\alpha _1} \cdots {\alpha _m}}^{(m)}(t - {t_1}, \cdots ,t - {t_m}) \cdot {E_{{\alpha _1}}}({t_1}) \cdots {E_{{\alpha _m}}}({t_m})} } \right\}}
\label{Eq-Pj_m}
\end{multline}

In frequency domain, it is:
\begin{multline}
\tilde P_j^{(m)}(\Omega )
= 2\pi {\varepsilon _0}\sum\limits_{{\alpha _1} \cdots {\alpha _m}} {\left\{ {\int_{ - \infty }^\infty  {d{\omega _1}} \cdots \int_{ - \infty }^\infty  {d{\omega _{m - 1}}} \tilde \chi _{j;{\alpha _1} \cdots {\alpha _m}}^{(m)}({\omega _1}, \cdots ,{\omega _{m - 1}},\Omega  - \sum\nolimits_1^{m - 1} {{\omega _i}} )} \right.}
\\
\left. { \times {{\tilde E}_{{\alpha _1}}}({\omega _1}) \cdots {{\tilde E}_{{\alpha _{m - 1}}}}({\omega _{m - 1}}){{\tilde E}_{{\alpha _m}}}(\Omega  - \sum\nolimits_1^{m - 1} {{\omega _i}} )} \right\}, m \ge 2
\label{Eq-Pj_m-FD}
\end{multline}
and ${\tilde P_j^{(1)}(\Omega ) = 2\pi {\varepsilon _0}\sum\limits_{{\alpha _1}} {\left\{ {\tilde \chi _{j;{\alpha _1}}^{(1)}(\Omega ){{\tilde E}_{{\alpha _1}}}(\Omega )} \right\}} }$. ${\tilde \chi _{j;{\alpha _1} \cdots {\alpha _m}}^{(m)}}$ corresponds to one component of the tensor ${\tilde \chi ^{(m)}}$, in which ${\tilde \chi _{j;{\alpha _1}}^{(1)}}$ is coming from the matrix ${\tilde \chi ^{(1)}}$.

Therefore, each component of the electric field ${\tilde E_j}$ has a wave equation which, in frequency domain, is written as:
\begin{equation}
{\nabla ^2}{\tilde E_j} + k_0^2(\omega ){\tilde E_j} + {\omega ^2}{\mu _0}{\tilde P_j} = 0;%
{{\tilde P}_j} = \tilde P_j^{(1)} + \tilde P_j^{(2)} +  \cdots  + \tilde P_j^{(m)}
\label{Eq-WaveEq-j}
\end{equation}

In particular, in uniaxial and biaxial crystals as well as cubic/isotropic materials, matrix ${\tilde \chi ^{(1)}}$ only has diagonal elements and therefore ${\tilde P_j^{(1)}(\Omega ) = 2\pi {\varepsilon _0}\tilde \chi _{j;j}^{(1)}(\Omega ){{\tilde E}_j}(\Omega )}$. we combine ${\tilde P_j^{(1)}}$ with ${k_0^2{{\tilde E}_j}}$ and update the wave equation as:

\begin{equation}
{\nabla ^2}{{\tilde E}_j} + k_j^2(\omega ){{\tilde E}_j} + {\omega ^2}{\mu _0}{{\tilde P}_{j,NL}} = 0;%
{{\tilde P}_{j,NL}} = \tilde P_j^{(2)} +  \cdots  + \tilde P_j^{(m)}
\label{Eq-WaveEq-j-uniaxial}
\end{equation}
where ${k_j^2(\omega ) = k_0^2(1 + 2\pi \tilde \chi _{j;j}^{(1)})}$ and the refractive index in the dimension ${j}$ is therefore defined as ${{n_j} = \sqrt {1 + 2\pi \tilde \chi _{j;j}^{(1)}} }$.

\section{1+1D NWEF}
\label{sec-3}
In the approximation of planer wave propagation, we neglect the spacial dynamics in the propagation of the electric field but focus on the temporal and spectral dynamics. the Laplace operator is therefore reduced to the only derivative with respect to propagation axis ${z}$ since the spacial dynamics is eliminated, i.e. ${{\nabla ^2} \to {\textstyle{{{\partial ^2}} \over {\partial {z^2}}}}}$. Therefore, \eqn{Eq-WaveEq-j-uniaxial} is reduced to a 1+1D wave equation:

\begin{equation}
\frac{{{\partial ^2}}}{{\partial {z^2}}}{{\tilde E}_j} + k_j^2(\omega ){{\tilde E}_j} + {\omega ^2}{\mu _0}{{\tilde P}_{j,NL}} = 0
\label{Eq-WaveEq-j-uniaxial-1D}
\end{equation}

By factoring out the fast dependence of the propagation coordinate from the electric field for all the frequencies, i.e. ${{\tilde E_j}(z,\omega ) = {\tilde A_j(z,\omega )}{e^{ - i{k_j(\omega )}z}}}$, we get:

\begin{equation}
\frac{{{\partial ^2}}}{{\partial {z^2}}}{{\tilde A}_j} - 2i{k_j}(\omega )\frac{\partial }{{\partial z}}{{\tilde A}_j} + {\omega ^2}{\mu _0}{{\tilde P}_{j,NL}}{e^{i{k_j}z}} = 0
\end{equation}

In the slowly varying spectral amplitude approximation (SVSAA) \cite{kolesik2012theory}, i.e. ${\left| {{\textstyle{\partial  \over {\partial z}}}{{\tilde A}_j}} \right| \ll \left| {{k_j}{{\tilde A}_j}} \right|}$, we have ${{\textstyle{{{\partial ^2}} \over {\partial {z^2}}}}{{\tilde A}_j} \ll {\textstyle{\partial  \over {\partial z}}}{k_j}{{\tilde A}_j}}$, which means the second-order derivative can be removed.

So, we have:

\begin{equation}
\frac{\partial }{{\partial z}}{{\tilde A}_j} =  - i\frac{{{\omega ^2}{\mu _0}}}{{2{k_j}(\omega )}}{{\tilde P}_{j,NL}}{e^{i{k_j}z}}
\end{equation}
and the reduced 1+1D Maxwell equation (also called forward Maxwell equation) regarding the electric field \cite{bullough1979solitons, husakou2001supercontinuum, conforti2010nonlinear}:

\begin{equation}
\frac{{\partial {{\tilde E}_j}}}{{\partial z}} + i{k_j}(\omega ){{\tilde E}_j} =  - i\frac{{{\omega ^2}{\mu _0}}}{{2{k_j}(\omega )}}{{\tilde P}_{j,NL}}
\label{Eq-FME-j-uniaxial}
\end{equation}

Among all the nonlinear induced polarizations, second-order and third order nonlinear induced polarizations are most concerned. For second-order nonlinear induced polarization, the response is always assumed instantaneous, giving rise to three-wave-mixing (TWM). The response function has ${\chi _{j;{\alpha _1}{\alpha _2}}^{(2)}({t_1},{t_2}) = \bar \chi _{j;{\alpha _1}{\alpha _2}}^{(2)}\delta ({t_1})\delta ({t_2})}$, where ${\bar \chi _{j;{\alpha _1}{\alpha _2}}^{(2)}}$ is a constant indicating the response intensity. Correspondently, in frequency domain, the susceptibility has ${\tilde \chi _{j;{\alpha _1}{\alpha _2}}^{(2)}({\omega _1},{\omega _2}) = {\textstyle{1 \over {{{(2\pi )}^2}}}}\bar \chi _{j;{\alpha _1}{\alpha _2}}^{(2)}}$ according to \eqn{Eq-chi_m}, i.e. constant for all the frequencies ${\omega _1}$ and ${\omega _2}$. Hence, the second-order nonlinear induced polarization ${P_j^{(2)}}$ has the same form as shown in \cite{conforti2010ultrabroadband,conforti2011modeling}:

\begin{equation}
P_j^{(2)}(t) = {\varepsilon _0}\sum\limits_{{\alpha _1}{\alpha _2}} {\left\{ {\bar \chi _{j;{\alpha _1}{\alpha _2}}^{(2)}{E_{{\alpha _1}}}{E_{{\alpha _2}}}} \right\}}
\label{Eq-Pj_2}
\end{equation}

\begin{equation}
\tilde P_j^{(2)}(\omega ) = {\varepsilon _0}\sum\limits_{{\alpha _1}{\alpha _2}} {\left\{ {\bar \chi _{j;{\alpha _1}{\alpha _2}}^{(2)}{{\tilde E}_{{\alpha _1}}}{\textstyle{ \otimes  \over {2\pi }}}{{\tilde E}_{{\alpha _2}}}} \right\}}  = {\varepsilon _0}\sum\limits_{{\alpha _1}{\alpha _2}} {\left\{ {\bar \chi _{j;{\alpha _1}{\alpha _2}}^{(2)}F\left[ {{E_{{\alpha _1}}}{E_{{\alpha _2}}}} \right]} \right\}}
\label{Eq-Pj_2-FD}
\end{equation}

As for third-order nonlinear induced polarization, the response is not fully instantaneous but consists of the instantaneous electronic Kerr response and a fraction of non-instantaneous vibrational Raman response. The response function is therefore written as ${\chi _{j;{\alpha _1}{\alpha _2}{\alpha _3}}^{(3)}({t_1},{t_2},{t_3}) = \bar \chi _{j;{\alpha _1}{\alpha _2}{\alpha _3}}^{(3)}R({t_1})\delta ({t_2} - {t_1})\delta ({t_3})}$, where ${R(t) = (1 - {f_R})\delta (t) + {f_R}{h_R}(t)}$ and ${\int_{ - \infty }^\infty  {h_R(t)dt}  = 1}$. ${f_R}$ indicates the amount of Raman fraction. ${h_R(t)}$ is the temporal Raman response function. The remaining instantaneous response is called electronic Kerr response which is the origin of the effects of self-phase modulation (SPM), cross-phase modulation (XPM) and four-wave-mixing (FWM). In frequency domain, the susceptibility is ${\tilde \chi _{j;{\alpha _1}{\alpha _2}{\alpha _3}}^{(3)}({\omega _1},{\omega _2},{\omega _3}) = {\textstyle{1 \over {{{(2\pi )}^3}}}}\bar \chi _{j;{\alpha _1}{\alpha _2}{\alpha _3}}^{(3)}\left[ {(1 - {f_R}) + {f_R}{{\tilde h}_R}({\omega _1} + {\omega _2})} \right]}$. Hence, the third-order nonlinear induced polarization ${P_j^{(3)}}$ has:

\begin{equation}
P_j^{(3)}(t) = {\varepsilon _0}\sum\limits_{{\alpha _1}{\alpha _2}{\alpha _3}} {\left\{ {\bar \chi _{j;{\alpha _1}{\alpha _2}{\alpha _3}}^{(3)}\left[ {(1 - {f_R}){E_{{\alpha _1}}}{E_{{\alpha _2}}}{E_{{\alpha _3}}} + {f_R}({h_R} \otimes ({E_{{\alpha _1}}}{E_{{\alpha _2}}})){E_{{\alpha _3}}}} \right]} \right\}}
\label{Eq-Pj_3}
\end{equation}

\begin{multline}
\tilde P_j^{(3)}(\omega ) = {\varepsilon _0}\sum\limits_{{\alpha _1}{\alpha _2}{\alpha _3}} {\left\{ {\bar \chi _{j;{\alpha _1}{\alpha _2}{\alpha _3}}^{(3)}\left[ {(1 - {f_R})({{\tilde E}_{{\alpha _1}}}{\textstyle{ \otimes  \over {2\pi }}}{{\tilde E}_{{\alpha _2}}}{\textstyle{ \otimes  \over {2\pi }}}{{\tilde E}_{{\alpha _3}}}) + {f_R}({\tilde h_R}({{\tilde E}_{{\alpha _1}}}{\textstyle{ \otimes  \over {2\pi }}}{{\tilde E}_{{\alpha _2}}})){\textstyle{ \otimes  \over {2\pi }}}{{\tilde E}_{{\alpha _3}}}} \right]} \right\}}  \\
= {\varepsilon _0}\sum\limits_{{\alpha _1}{\alpha _2}{\alpha _3}} {\left\{ {\bar \chi _{j;{\alpha _1}{\alpha _2}{\alpha _3}}^{(3)}\left[ {(1 - {f_R})F\left[ {{E_{{\alpha _1}}}{E_{{\alpha _2}}}{E_{{\alpha _3}}}} \right] + {f_R}F\left[ {{E_{{\alpha _3}}}{F^{ - 1}}\left[ {{{\tilde h}_R}F\left[ {{E_{{\alpha _1}}}{E_{{\alpha _2}}}} \right]} \right]} \right]} \right]} \right\}}
\label{Eq-Pj_3-FD}
\end{multline}

Finally, by substituting the nonlinear induced polarizations in \eqn{Eq-FME-j-uniaxial} with their interpretations \eqn{Eq-Pj_2-FD} and \eqn{Eq-Pj_3-FD}, we get the 1+1D NWEF:

\begin{multline}
\frac{{\partial {{\tilde E}_j}}}{{\partial z}} + i{k_j}(\omega ){{\tilde E}_j} =  - i\frac{{{\omega ^2}}}{{2{c^2}{k_j}(\omega )}} \sum\limits_{{\alpha _1}{\alpha _2}} {\left( {\bar \chi _{j;{\alpha _1}{\alpha _2}}^{(2)}F\left[ {{E_{{\alpha _1}}}{E_{{\alpha _2}}}} \right]} \right)} \\
 - i\frac{{{\omega ^2}}}{{2{c^2}{k_j}(\omega )}} \sum\limits_{{\alpha _1}{\alpha _2}{\alpha _3}} {\left\{ {\bar \chi _{j;{\alpha _1}{\alpha _2}{\alpha _3}}^{(3)}\left[ {\left( {1 - {f_R}} \right)F\left[ {{E_{{\alpha _1}}}{E_{{\alpha _2}}}{E_{{\alpha _3}}}} \right] + {f_R}F\left[ {{E_{{\alpha _3}}}{F^{ - 1}}\left[ {{{\tilde h}_R}F\left[ {E_{{\alpha _1}}E_{{\alpha _2}}} \right]} \right]} \right]} \right]} \right\}}
\label{Eq-NWEF-1D}
\end{multline}

\Eqn{Eq-NWEF-1D} can numerically be solved by split-step Fourier method together with Runge-Kutta method. It is noted that the frequency domain has a range ${( - \infty , + \infty )}$ and the contents of ${k_j(\omega )}$ in negative frequencies are required. According to the causality of all the induced polarizations, ${k_j(\omega )}$ shows a property of complex conjugate, i.e. its contents in negative frequencies are linked to what in positive frequencies.

\section{NWEF in waveguide structure}
\label{sec-4}
In a waveguide structure, the spacial dynamics of the electric field is always dominated by the eigen modes in the waveguide, which determine the spacial distribution as well as the propagation constant  of the electric field. Since the orthogonality between any of the two eigen modes, we redefine the dimension mark ${j}$ in the wave equation \eqn{Eq-WaveEq-j-uniaxial} to be the mode mark. Moreover, the electric field is redefined as \cite{phillips2011supercontinuum} ${{\tilde E_j}(x,y,z,\omega ) = \tilde B_j(x,y,\omega )\tilde A_j^\varphi (z,\omega ) = \tilde B_j(x,y,\omega )\tilde A_j(z,\omega ){e^{ - i{\beta _j}(\omega )z}}}$, where ${\tilde B_j}$ is the eigen mode distribution and ${\beta _j}$ indicates the mode propagation constant. Now \eqn{Eq-WaveEq-j-uniaxial} can be expanded as:

\begin{equation}
{{\tilde A}_j}{e^{ - i{\beta _j}(\omega )z}} \left( {\frac{{{\partial ^2}}}{{\partial {x^2}}} + \frac{{{\partial ^2}}}{{\partial {y^2}}} + k_j^2 - \beta _j^2} \right){{\tilde B}_j} + {{\tilde B}_j}{e^{ - i{\beta _j}(\omega )z}} \left( {\frac{{{\partial ^2}}}{{\partial {z^2}}} - i2{\beta _j}\frac{\partial }{{\partial z}}} \right){{\tilde A}_j} + {\omega ^2}{\mu _0}{{\tilde P}_{j,NL}} = 0
\end{equation}

Remember all the eigen modes of the waveguide have ${({\textstyle{{{\partial ^2}} \over {\partial {x^2}}}} + {\textstyle{{{\partial ^2}} \over {\partial {y^2}}}} + k_j^2 - \beta _j^2) = 0}$ and using SVSAA, we get the reduced Maxwell equation for waveguide structure:

\begin{equation}
\frac{{\partial {{\tilde E}_j}}}{{\partial z}} + i{\beta _j}(\omega ){{\tilde E}_j} =  - i\frac{{{\omega ^2}{\mu _0}}}{{2{\beta _j}(\omega )}}{{\tilde P}_{j,NL}}
\label{Eq-FME-j-uniaxial-waveguide}
\end{equation}

Compared with the 1+1D forward Maxwell equation \eqn{Eq-FME-j-uniaxial}, the only difference in the above expression is the replacement of the spacial propagation constant ${k_j}$ by mode propagation constant ${\beta _j}$. However, it should be noticed that not only the dispersion characteristics but the nonlinear induced polarizations are all revised by employing the waveguide, which will be revealed later.

Instead of focusing on the space-related electric field, now we concentrate on the spectral related amplitude ${\tilde A_j}$ and get its dynamics by making spacial integral on both sides of \eqn{Eq-FME-j-uniaxial-waveguide}, i.e.:

\begin{equation}
\iint_\infty  {dxdy\tilde B_j^*\left( {\frac{\partial }{{\partial z}} + i{\beta _j}} \right){{\tilde E}_j}} =  - i\frac{{{\omega ^2}{\mu _0}}}{{2{\beta _j}}}\iint_\infty  {dxdy\tilde B_j^*{{\tilde P}_{j,NL}}}
\end{equation}

\begin{equation}
\frac{{\partial \tilde A_j^\varphi }}{{\partial z}} + i{\beta _j}(\omega )\tilde A_j^\varphi  =  - i\frac{{{\omega ^2}{\mu _0}}}{{2{{\tilde g}_j}(\omega ){\beta _j}(\omega )}}\iint_\infty  {dxdy\tilde B_j^*{{\tilde P}_{j,NL}}}
\label{Eq-FME-j-uniaxial-waveguide-amp}, {\iint_\infty  {dxdy{{\left| {{{\tilde B}_j}} \right|}^2}} = {\tilde g_j}(\omega )}
\end{equation}

Analogously, we should interpret the nonlinear induced polarizations in \eqn{Eq-FME-j-uniaxial-waveguide-amp} in which the nonlinear susceptibilities are now space-dependent. For second-order nonlinear induced polarization, we employ \eqn{Eq-Pj_m-FD} and expand it with the definition of the electric field, getting:

\begin{multline}
\iint_\infty  {dxdy\tilde B_j^*\tilde P_j^{(2)}} = 2\pi \varepsilon _0\sum\limits_{\alpha _1\alpha _2}{\int\limits_{ - \infty }^\infty  {\{ d{\omega _1}\tilde A_{{\alpha _1}}^\varphi ({\omega _1})\tilde A_{{\alpha _2}}^\varphi (\Omega  - {\omega _1})}} \\
 \times \iint_\infty  {dxdy\tilde \chi _{j;{\alpha _1}{\alpha _2}}^{(2)}(x,y,{\omega _1},\Omega  - {\omega _1})\tilde B_j^*(x,y,\Omega ){{\tilde B}_{{\alpha _1}}}(x,y,{\omega _1}){{\tilde B}_{{\alpha _2}}}(x,y,\Omega  - {\omega _1})}\}
\end{multline}
in which we can define a spacial integral factor as Phillips et al. did in \cite{phillips2011supercontinuum}:

\begin{multline}
{{\tilde \Theta }_{j;{\alpha _1}{\alpha _2}}^{(2)}}({\omega _1},{\omega _2}) = \iint_\infty  {dxdy\tilde \chi _{j;{\alpha _1}{\alpha _2}}^{(2)}(x,y,{\omega _1},{\omega _2})\tilde B_j^*(x,y,{\omega _1} + {\omega _2}){{\tilde B}_{{\alpha _1}}}(x,y,{\omega _1}){{\tilde B}_{{\alpha _2}}}(x,y,{\omega _2})} \\
= \frac{1}{{{{(2\pi )}^2}}}\iint_\infty  {dxdy\bar \chi _{j;{\alpha _1}{\alpha _2}}^{(2)}(x,y)\tilde B_j^*(x,y,{\omega _1} + {\omega _2}){{\tilde B}_{{\alpha _1}}}(x,y,{\omega _1}){{\tilde B}_{{\alpha _2}}}(x,y,{\omega _2})}
\end{multline}
and simplify the expression of second-order nonlinear induced polarization as:

\begin{equation}
\iint_\infty  {dxdy\tilde B_j^*\tilde P_j^{(2)}} = 2\pi \varepsilon _0\sum\limits_{\alpha _1\alpha _2}{\int\limits_{ - \infty }^\infty  {d{\omega _1}\tilde A_{{\alpha _1}}^\varphi ({\omega _1})\tilde A_{{\alpha _2}}^\varphi (\Omega  - {\omega _1}){{\tilde \Theta }_{j;{\alpha _1}{\alpha _2}}^{(2)}}({\omega _1},\Omega  - {\omega _1})}}
\end{equation}

Now the spacial integral factor ${\tilde \Theta _{j;\alpha _1\alpha _2}^{(2)}}$ plays the role as the second-order susceptibility, determining the intensity of the nonlinear induced polarization. Moreover, it can be assumed that the variation of the mode distribution ${\tilde B_j}$ with respect to frequency is slow enough compared with that of ${\tilde A_{j}^\varphi }$, or named slowly varying mode distribution approximation (SVMDA), therefore the spacial integral factor is considered constant ${{{\tilde \Theta }_{j;{\alpha _1}{\alpha _2}}^{(2)}}({\omega _1},{\omega _2}) \approx \tfrac{1}{{{{(2\pi )}^2}}}{{\bar \Theta }_{j;{\alpha _1}{\alpha _2}}^{(2)}}}$ and the second-order induced polarization is further simplified to:

\begin{equation}
\iint_\infty  {dxdy\tilde B_j^*\tilde P_j^{(2)}} = {\varepsilon _0}\sum\limits_{{\alpha _1}{\alpha _2}} {\left\{ {{{\bar \Theta }_{j;{\alpha _1}{\alpha _2}}^{(2)}}\tilde A_{{\alpha _1}}^\varphi \tfrac{ \otimes }{{2\pi }}\tilde A_{{\alpha _2}}^\varphi } \right\}}  = {\varepsilon _0}\sum\limits_{{\alpha _1}{\alpha _2}} {\left\{ {{{\bar \Theta }_{j;{\alpha _1}{\alpha _2}}^{(2)}}F\left[ {A_{{\alpha _1}}^\varphi A_{{\alpha _2}}^\varphi } \right]} \right\}}
\label{Eq-Pj_2-FD-waveguide}
\end{equation}

Analogously, for third-order nonlinear induced polarization, we have:

\begin{equation}
\iint_\infty  {dxdy\tilde B_j^*\tilde P_j^{(3)}} = {\varepsilon _0}\sum\limits_{{\alpha _1}{\alpha _2}{\alpha _3}} {\left\{ {\bar \Theta _{j;{\alpha _1}{\alpha _2}{\alpha _3}}^{(3)}\left[ {(1 - {f_R})F\left[ {A_{{\alpha _1}}^\varphi A_{{\alpha _2}}^\varphi A_{{\alpha _3}}^\varphi } \right] + {f_R}F\left[ {A_{{\alpha _3}}^\varphi {F^{ - 1}}\left[ {{{\tilde h}_R}F\left[ {A_{{\alpha _1}}^\varphi A_{{\alpha _2}}^\varphi } \right]} \right]} \right]} \right]} \right\}}
\label{Eq-Pj_3-FD-waveguide}
\end{equation}

The spacial integral factor playing a role as the third-order susceptibility is:

\begin{multline}
\tilde \Theta _{j;{\alpha _1}{\alpha _2}{\alpha _3}}^{(3)}({\omega _1},{\omega _2},{\omega _3}) \\
= \iint_\infty  {dxdy\tilde \chi _{j;{\alpha _1}{\alpha _2}{\alpha _3}}^{(3)}(x,y,{\omega _1},{\omega _2},{\omega _3})\tilde B_j^*(x,y,\sum\nolimits_n {{\omega _n}} ){{\tilde B}_{{\alpha _1}}}(x,y,{\omega _1}){{\tilde B}_{{\alpha _2}}}(x,y,{\omega _2}){{\tilde B}_{{\alpha _3}}}(x,y,{\omega _3})} \\
= \frac{{\left[ {1 - {f_R} + {f_R}{{\tilde h}_R}({\omega _1} + {\omega _2})} \right]}}{{{{(2\pi )}^3}}}\iint_\infty  {dxdy\bar \chi _{j;{\alpha _1}{\alpha _2}{\alpha _3}}^{(3)}\tilde B_j^*(x,y,\sum\nolimits_n {{\omega _n}} ){{\tilde B}_{{\alpha _1}}}(x,y,{\omega _1}){{\tilde B}_{{\alpha _2}}}(x,y,{\omega _2}){{\tilde B}_{{\alpha _3}}}(x,y,{\omega _3})} \\
\approx \frac{1}{{{{(2\pi )}^3}}}\bar \Theta _{j;{\alpha _1}{\alpha _2}{\alpha _3}}^{(3)}\left[ {1 - {f_R} + {f_R}{{\tilde h}_R}({\omega _1} + {\omega _2})} \right]
\end{multline}

Finally, by substituting the nonlinear induced polarizations in \eqn{Eq-FME-j-uniaxial-waveguide-amp} with their interpretations \eqn{Eq-Pj_2-FD-waveguide} and \eqn{Eq-Pj_3-FD-waveguide}, we get the NWEF in the waveguide structure:

\begin{multline}
\frac{{\partial \tilde A_j^\varphi }}{{\partial z}} + i{\beta _j}(\omega )\tilde A_j^\varphi = - i\frac{{{\omega ^2}}}{{2c^2{{\tilde g}_j}(\omega ){\beta _j}(\omega )}} \sum\limits_{{\alpha _1}{\alpha _2}} {\left\{ {{{\bar \Theta }_{j;{\alpha _1}{\alpha _2}}^{(2)}}F\left[ {A_{{\alpha _1}}^\varphi A_{{\alpha _2}}^\varphi } \right]} \right\}}  \\
- i\frac{{{\omega ^2}}}{{2c^2{{\tilde g}_j}(\omega ){\beta _j}(\omega )}}\sum\limits_{{\alpha _1}{\alpha _2}{\alpha _3}} {\left\{ {\bar \Theta _{j;{\alpha _1}{\alpha _2}{\alpha _3}}^{(3)}\left[ {(1 - {f_R})F\left[ {A_{{\alpha _1}}^\varphi A_{{\alpha _2}}^\varphi A_{{\alpha _3}}^\varphi } \right] + {f_R}F\left[ {A_{{\alpha _3}}^\varphi {F^{ - 1}}\left[ {{{\tilde h}_R}F\left[ {A_{{\alpha _1}}^\varphi A_{{\alpha _2}}^\varphi } \right]} \right]} \right]} \right]} \right\}}
\label{}
\end{multline}

\bibliographystyle{osajnl}
\bibliography{ref}




\end{document}